\documentclass[prl,aps, amsfonts, nofootinbib]{revtex4}

\usepackage{epsf}

\def\CPT{$\chi${\rm PT}}
\def\CPTs{$\chi${\rm PT }}

\newcommand{\beq}{\begin{equation}}
\newcommand{\eeq}{\end{equation}}
\newcommand{\bea}{\begin{eqnarray}}
\newcommand{\eea}{\end{eqnarray}}
\newcommand{\eps}{\epsilon}

\newcommand{\half}{{\textstyle\frac{1}{2}}}

\newcommand{\nn}{\nonumber}
\newcommand{\benn}{\begin{displaymath}}
\newcommand{\eenn}{\end{displaymath}}
\newcommand{\ket}[1]{| #1 \rangle}                     
\newcommand{\bra}[1]{\langle #1 \, |}                  
\newcommand{\no}{\frac{2\pi n_0}{\beta}}
\newcommand{\nv}{\frac{2\pi \vec{n}}{L}}
\newcommand{\tr}{{\rm Tr}}
\newcommand{\A}{{\mathcal A}}
\newcommand{\V}{{\mathcal V}}
\newcommand{\M}{{\mathcal M}}
\newcommand{\Op}{{\mathcal O}}
\newcommand{\hinv}{h^{-1}}

\begin{document}
\flushleft{TUM-T39-04-10, LA-UR-04-4810, LBNL-55716, hep-lat/0407009}
\flushleft{1st July 2004}
\flushleft{Revised version 1st January 2005, re-revised version 10th March
  2005. Accepted by Phys.~Rev.~\textbf{D}.}

\title{A nucleon in a tiny box}

\author{Paulo F.~Bedaque\footnote{{\tt pfbedaque@lbl.gov}}}
\affiliation{Lawrence-Berkeley Laboratory, Berkeley, CA 94720, USA}
\author{Harald W.~Grie\3hammer\footnote{{\tt hgrie@ph.tum.de }}}
\affiliation{Institut f\"ur Theoretische Physik (T39), Technische
  Universit\"at M\"unchen, D-85747 Garching, Germany} \author{Gautam
  Rupak\footnote{{\tt grupak@lanl.gov}}} \affiliation{Los Alamos National
  Laboratory, Los Alamos, NM 87545, USA}

\begin{abstract}
  We use Chiral Perturbation Theory  to compute the nucleon
  mass-shift due to finite volume and temperature effects. Our results are
  valid up to next-to-leading order in the ``$\eps$-r{\'e}gime" ($mL\sim
  m\beta\ll 1$) as well as in the ``$p$-r{\'e}gime" ($mL\sim m\beta\gg 1$).
  Based on the two leading orders, we discuss the convergence of the expansion
  as a function of the lattice size and quark masses.  This result can be used
  to extrapolate lattice results obtained from lattice sizes smaller than the
  pion cloud, avoiding the numerical simulation of physics under theoretical
  control. An extraction of the low-energy coefficient $c_3$ of the chiral
  Lagrangean from lattice simulations at small volumes and a ``magic'' ratio
  $\beta=1.22262 L$ might be possible.
\end{abstract}
\maketitle

Lattice QCD simulations are necessarily performed in finite boxes.
Finite-size effects are controlled by the parameter $mL$, where $L$ is the
lattice size and $m$ the mass of the lightest particle, in QCD, the pion.
Physical results can be obtained in the limit $mL\gg 1$. As the pion masses
achieved in simulations approach the physical value it becomes harder to
fulfill this condition.  However, most of the configurations in large lattices
describe pions traveling at large distances of the order of $L$.  Since the
physics of these soft-pions is strongly constrained by chiral symmetry, strong
theoretical control over them makes their numerical simulation unnecessary.
One can thus obtain physical results by simulating in smaller lattices and
using Chiral Perturbation Theory (\CPT) or some other relevant effective
theory to include the soft-pion physics cut off by the box size and
extrapolate the results to the infinite volume limit.  Another way to describe
the same procedure gives added insight: The low-energy physics in the infinite
and finite volume are described by the same effective theory with the same
low-energy constants, since the values of these constants encapsulate
short-distance physics that is not modified by finite-volume effects. The
comparison of finite volume lattice results with the effective theory
prediction allows one therefore to determine the value of some of the low
energy constants. Those, in turn, can be used to determine physical
observables in the infinite-volume limit.

This general procedure has been carried out in the r{\'e}gime $mL\gg 1$, where
standard~\CPTs can be applied, to a variety of one nucleon observables, see
e.g.~\cite{finite_volume}. However, it is for $mL\ll 1$ (in the so-called
$\eps$-r{\'e}gime \cite{leutwyler_epsilon}) that the programme described above
is fully realized. For such small boxes, most of the pion cloud surrounding a
baryon is excluded, and we are left with a bare nucleon.  There are some
modifications to the usual~\CPTs power counting in this r{\'e}gime. The first
and obvious one is that the momenta are quantized in units of $2\pi/L$. More
importantly, the pion zero mode fluctuations are not suppressed, become
non-perturbative and need to be treated exactly \cite{leutwyler_epsilon}. They
reduce the value of the chiral condensate and make the chiral condensate
disappear altogether in the chiral limit. This is to be expected since there
is no chiral symmetry breaking at finite volumes.  Recently, the
$\eps$-r{\'e}gime in the meson sector and its relevance to lattice QCD have
been assessed in a number of papers \cite{epsilon_mesons}. In the present
work, we extend the idea to the one-baryon sector.  Convergence in the
baryonic sector is typically worse than in the mesonic sector, as it receives
contributions at every order in $p/(4\pi f)$, unlike the meson sector case
where the expansion parameter is $(p/(4\pi f))^2$. We address this issue by
comparing the sizes of leading and next-to-leading order contributions in a
calculation of the nucleon mass.


We consider one nucleon in a small box of size $\beta\times L^3$ for
$2\pi/(4\pi f)\alt\beta,L \alt 2\pi/m$, the ``$\epsilon$-r{\'e}gime". $L$ is
the size of the spatial directions, $\beta$ the temporal extend of the box,
namely the inverse temperature. In this r{\'e}gime,~\CPTs is valid, except
that the relative counting between $p$ and $m$ is changed.  Instead of the
usual counting $1/L,1/\beta\sim m\sim p$ ($p$-counting), we use
$1/L,1/\beta\sim m_q\sim \sqrt{m}\sim\epsilon$, hence the name
``$\epsilon$-r{\'e}gime"~\cite{leutwyler_epsilon}. For small boxes, the first
non-zero pion mode has a momentum $p=2\pi/L \agt \Delta$, so we include the
$\Delta(1232)$ as explicit degree of freedom, counting, in the
$\eps$-r\'egime, $\Delta\sim m \sim \epsilon^2$.

\section{The $\eps$ expansion in the baryon sector} Low-energy properties
($Q\sim 1/L,1/\beta$) of the system are described by the effective Euclidean
Lagrangean
\begin{eqnarray}
  {\mathcal L}&=& {\mathcal L}_\pi+{\mathcal L}_N+{\mathcal L}_\Delta \nn,\\
  {\mathcal L}_\pi&=&
  f^2 \tr\A_\mu\A_\mu -\frac{Bf^2}{2}\tr(\xi_R^\dagger\M\xi_L+\xi_L^\dagger\M^\dagger\xi_R)+\cdots\nn,\\
  {\mathcal L}_N&=&N^\dagger D_0 N - g_A N^\dagger \vec{\sigma}\cdot\vec{\A} N
  +N^\dagger [-\frac{\vec{D}^2}{2M}+\frac{g_A}{2M}\{\vec{\sigma}\cdot\vec{D},
  \A_0\}
  -2B c_1 \tr( \xi_R^\dagger\M\xi_L+\xi_L^\dagger\M^\dagger\xi_R )\nonumber\\
  && +4(c_2-\frac{g_A^2}{8M})\A_0^2+4c_3 \A_\mu\A_\mu
  -(c_4+\frac{1}{4M})2i\epsilon^{ijk} \A_i\A_j\sigma^k +\cdots)  ]N \nn,\\
  {\mathcal L}_\Delta&=&-\Delta^{\dagger
    iA}(D_0+\Delta-\frac{\vec{D}^2}{2M})\Delta^{iA}
  +g_{N\Delta}\Delta^{\dagger iA}(w^A_i N+ \mathrm{H.c.})  +\cdots,
  \label{eq:lagr}
\end{eqnarray}
where we list only the terms pertinent to our calculation.  The pion decay
constant is $f=92.4$ MeV, $\xi_L, \xi_R$ are $SU(2)$ matrices parameterizing
the chiral $SU_L(2)\times SU_R(2)$ group and $\M={\rm diag}(m_q,m_q)$ is the
quark mass matrix in the isospin limit (the precise conventions used can be
found in Appendix A). The values of the other low-energy constants will be
given when we discuss our results. The Goldstone bosons belong to the coset
space $[SU_L(2)\times SU_R(2)]/SU_{L+R}(2)$, and we are free to choose an
arbitrary member to be the representative of each coset (``fix the gauge").
Instead of the usual choice
$\xi_L=\xi_R^\dagger=\xi=e^{\frac{i\pi\cdot\tau}{2f}}$, we use the choice made
in background field calculations
\begin{eqnarray}
\xi_L&=&u_0 e^{\frac{i\pi\cdot\tau}{2f}}\nonumber,\\
\xi_R&=&u_0^\dagger e^{-\frac{i\pi\cdot\tau}{2f}},
\end{eqnarray}
where $u_0$ is a space-time independent field and $\pi(x)$ does not contain
zero-modes:
\begin{equation}
\pi(x)=\sum\limits_{n_\mu\neq (0,\vec{0})}\pi_n
\ e^{i\no t + i \nv\cdot\vec{x}}.
\end{equation}
The rationale to separate zero- and non-zero modes is that, as we will see
below, the zero modes obey a different power counting from the non-zero ones
at small volumes where chiral symmetry is partially restored.

The background field $u_0$ only appears in those terms of the action which
include quark masses. This can be easily seen by noticing that a non-trivial
background $u_0$ corresponds to a chiral rotation of the vacuum one expands
around. In the absence of quark masses, all such vacua are equivalent, so the
physics of the Goldstone bosons is the same. The terms which do however depend
on the quark masses are in the isospin limit with
$\mathcal{Re}\tr(A)=\half\tr(A+A^\dagger)$:
\begin{equation}\label{zeromode_integrand}
  - m_qBf^2 \mathcal{Re}\tr(u_0^2 e^{\frac{i\pi\cdot\tau}{f}})
  -4 m_qB c_1 N^\dagger{\mathcal Re}\tr(u_0^2e^{\frac{i\pi\cdot\tau}{f}}) N .
\end{equation}
At leading order, $m^2=2Bm_q$ is the pion mass in the infinite volume limit.

\begin{figure}[!htbp]
  \centerline{{\epsfxsize=2.5in \epsfbox{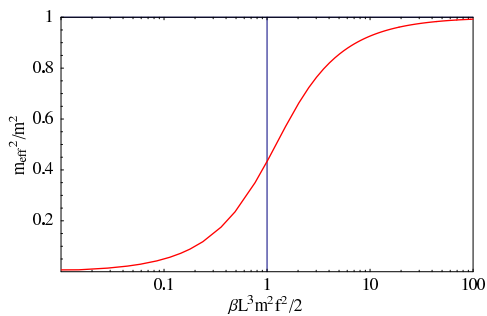}}}
\noindent
\caption{The effective pion mass $m_\mathrm{eff}$ as function of $\beta L^3
  m^2f^2/2$, Eq.~(\ref{meff}). }
\label{fig:pionmass}
\end{figure}

We can now estimate the different terms of the Lagrangean. The typical
fluctuations of the non-zero modes $\pi(x)$ are of the order $\pi(x)\sim \eps$
since, for larger values of $\pi(x)$, the kinetic term is much larger than one
and suppresses their contribution to the path integral (we can estimate the
size of the kinetic term as $1/\epsilon^4$ coming from the volume integral,
$\epsilon^2$ from the two derivatives and $\epsilon^2$ from the pion fields).
A similar argument implies $N\sim \epsilon^{3/2}$. However, as observed by
Gasser and Leutwyler \cite{leutwyler_epsilon}, the zero-mode $u_0$ is of order
$\eps^0$.  We can conclude that by noticing that the coefficient of the first
term of Eq.(\ref{zeromode_integrand}) is of order $\sim m^2 f^2\beta L^3\sim
\eps^0 $.  Because the zero-mode is not suppressed, it has to be treated
exactly. This is related to the restoration of chiral symmetry at finite
temperatures and volumes. In small boxes the zero-mode fluctuates over the
whole group manifold, in contradistinction to the infinite volume limit in
which the zero-mode makes only small fluctuations around a preferred vacuum
direction.  As shown in \cite{leutwyler_epsilon} the integration over the
zero-mode can be performed as follows.  The part of the partition function
which contains $u_0$ can be written as
\begin{eqnarray}
  Z[N,\Delta]&=&\int [Du_0]\; \exp\left[\int
  d^4x\left(\frac{m^2f^2}{2}+2m^2c_1N^\dagger(x)N(x) \right) \left({\mathcal
      Re}\tr(u_0^2)\left(1-\frac{\pi^2}{2f^2}\right)+ \frac{1}{f^2} {\mathcal
      Re}\tr(u^2i\pi^A\tau^A)
    +\cdots\right)\right]\nn\\
&\approx&\int [Du_0]\; e^{s{\mathcal Re}\tr(u_0^2)}
\bigg[1+{\mathcal Re}\tr(u_0^2)\bigg(2m^2c_1\int d^4x\;N^\dagger N
\left(1-\frac{\pi^2}{2f^2}\right)
-\frac{m^2}{4}\int d^4x \;\pi^2\bigg)\bigg]\nn\\
&\approx& X(s)\; \exp[
-\frac{X^\prime(s)}{2X(s)} \int d^4x\;
  \left(-4m^2c_1N^\dagger(x)N(x)
    \left(1-\frac{\pi^2}{2f^2}\right)+\frac{m^2}{2} \pi^2
  \right)]
\end{eqnarray}
where we dropped higher orders in the pion-fields, $s=Bf^2m_q\beta
L^3=m^2f^2\beta L^3/2$ and
\begin{equation}
  X(s)=\int_{SU(2)}\!\!\!\!\!\! [Du_0^2] \ \ 
e^{s{\mathcal Re}\tr(u_0^2)}=\frac{I_1(2s)}{s} ,
\end{equation}
with $I_1(x)$ a modified Bessel function.  The integration over the zero-mode
performed above renormalizes the nucleon mass (adding a term proportional to
$c_1$ of order $\eps^4$ to it) and the pion mass (by a term of order
$\eps^0$), as well as the non-derivative couplings:
\begin{eqnarray}
  M&\to&M-4m^2 c_1\;
\frac{X^\prime(\beta L^3m^2f^2/2)}{2X(\beta L^3m^2f^2/2)},\nn\\
m^2&\to&m_\mathrm{eff}^2=\underbrace{2m_qB}_{m^2} \frac{X^\prime(\beta
  L^3m^2f^2/2)}{2X(\beta L^3m^2f^2/2)}\label{meff}.
\end{eqnarray} The effective pion mass is shown in Fig.~\ref{fig:pionmass}.
In the limit $s\to\infty$, one retrieves with
$X^\prime(\infty)/(2X(\infty))=1$, the well-known infinite-volume results.
\bigskip
Notice that the shift of the nucleon and pion masses due to the zero modes
comes just from quenching the chiral condensate in the finite volume:
\begin{equation}
  \frac{\ \ \ \ \bra{0}\bar q q\ket{0}_{\beta,L}}{\bra{0}\bar q q\ket{0}}
=\frac{X'(s)}{2X(s)}.
\end{equation}
The total partition function of the system is finally to the order considered
\begin{equation}
  Z=\int[DN]\;[D\Delta]\;e^{-\mathcal{S}^\prime}\;Z[N,\Delta],
\end{equation}
with $\mathcal{S}^\prime$ the part of the action (\ref{eq:lagr}) which is
independent of zero-modes.

\section{Nucleon mass}
 
\begin{figure}[!htbp]
  \centerline{{\epsfxsize=4.1in \epsfbox{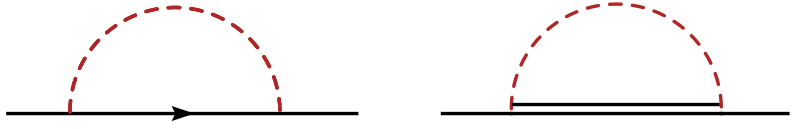}}}
\noindent
\caption{Leading-order contributions to the nucleon mass.}
\label{fig:lo}
\end{figure}

The shift in the nucleon mass due to finite volume effects is given at leading
order [${\mathcal O}(\eps^3)$] by the two one-loop diagrams of
Fig.~\ref{fig:lo}.  We find for the first diagram:
\begin{eqnarray}
  M_a^{(3)}(\beta,L)&=&-\frac{3g_A^2}{4f^2}\frac{1}{\beta
  L^3}\sum_{n_\mu\neq 0}
\frac{i}{\omega+\no}\frac{(\nv)^2}{(\no)^2+(\nv)^2+m_\mathrm{eff}^2}\nn\\
&\stackrel{\omega\rightarrow 0 }{\longrightarrow}&
-i\frac{3g_A^2}{4f^2}{\mathbb A}(\Delta= 0,m_\mathrm{eff}).
\end{eqnarray}
The second diagram is
\begin{eqnarray}
  M_b^{(3)}(\beta,L)&=&-\frac{4g_{N\Delta}^2}{3f^2}\sum_{n_\mu\neq
  0}\frac{i}{\omega+\no+i\Delta}
\frac{(\nv)^2}{(\no)^2+(\nv)^2+m^2_\mathrm{eff}}\nn\\
&\stackrel{\omega\rightarrow 0 }{\longrightarrow}&
-i\frac{4g_{N\Delta}^2}{3f^2}{\mathbb A}(\Delta,m_\mathrm{eff}).
\end{eqnarray}
Because $mL, m\beta, \Delta L$ and $\Delta\beta$ are all of order $\eps$ in
the $\eps$ expansion, the $m$ and $\Delta$ contribution to these graphs are of
order ${\mathcal O}(\eps^5)$, so that $m$ (and with that of course
$m_\mathrm{eff}$) and $\Delta$ can be dropped from the expressions above.

\begin{figure}[!htbp]
  \centerline{{\epsfxsize=4.1in \epsfbox{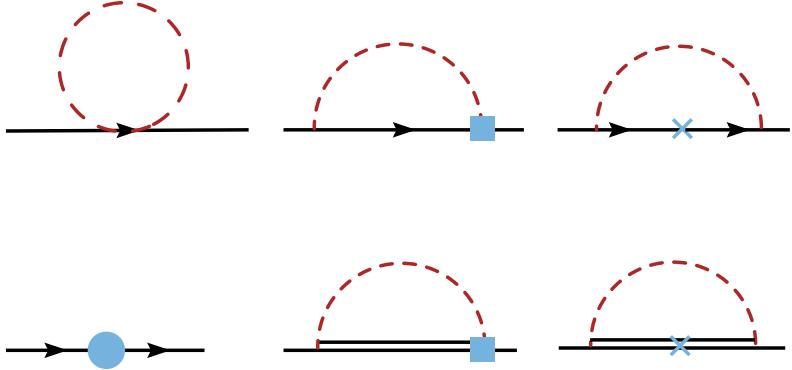}}}
\noindent
\caption{Next-to-leading order diagrams for the nucleon mass. The square
  vertex represents a vertex suppressed by $\eps$, the cross a kinetic energy
  insertion, and the circle the zero-mode mass contribution. The dashed, full
  and double lines represent a pion, nucleon and $\Delta$ propagator,
  respectively.}
\label{fig:nlo}
\end{figure}

\begin{figure}[!htbp]
  \centerline{{\epsfxsize=2.6in \epsfbox{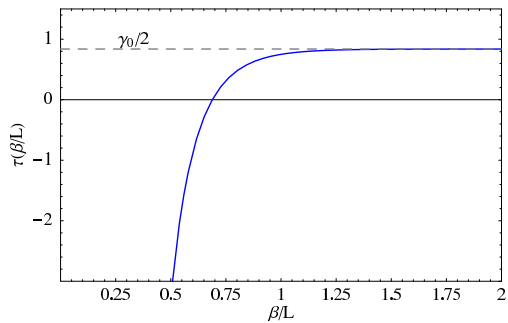}}}
\noindent
\caption{ $\tau(\beta/L)$ as a function of the box asymmetry $\beta/L$.}
\label{fig:tau}
\end{figure}

The first truly specific feature of the ``$\eps$-r{\'e}gime" appears at order
$\eps^4$, because the nucleon mass receives contributions from the zero modes
computed above in Eq.~(\ref{meff}) in addition to the graphs shown in
Fig.~\ref{fig:nlo}. The first graph leads to
\begin{eqnarray}
  M_a^{(4)}(\beta,L)&=& -\frac{3\bar c_2+3c_3}{f^2}\frac{1}{\beta L^3}
\sum_{n_\mu\neq 0}\frac{(\no)^2}{(\no)^2+(\nv)^2+m_\mathrm{eff}^2}
-\frac{3c_3}{f^2}\frac{1}{\beta L^3}
\sum_{n_\mu\neq 0}\frac{(\nv)^2}{(\no)^2+(\nv)^2+m_\mathrm{eff}^2}\nn\\
&=&-\frac{3\bar c_2+3c_3}{f^2} {\mathbb C}(m_\mathrm{eff})-
\frac{3c_3}{f^2}{\mathbb D}(m _\mathrm{eff}), 
\end{eqnarray}
with $\bar{c_2}=c_2-\frac{g_A2}{8M}$.  The second and fifth graph vanishes.
The third one gives
\begin{eqnarray}
  M_c^{(4)}(\beta,L)&=&\frac{3g_A^2}{8Mf^2}\frac{1}{\beta
  L^3}\sum_{n_\mu\neq 0}
\left(\frac{i}{\omega+\no}\right)^2\frac{(\nv)^4}{(\no)^2+(\nv)^2+m_\mathrm{eff}^2}
\nn\\
&=&-\frac{3g_A^2}{8Mf^2}{\mathbb B}(\Delta= 0,m_\mathrm{eff}).
\end{eqnarray}
The fourth graph is the non-perturbative contribution computed before in
Eq.~(\ref{meff}) as
\begin{equation}
  M_d^{(4)}(\beta,L) = -2m^2 c_1 \frac{X'(m^2f^2\beta L^3/2)} {X
  (m^2f^2\beta L^3/2)},
\end{equation}
and the last one contributes as
\begin{eqnarray}
  M_f^{(4)}(\beta,L)&=&\frac{2g_{N\Delta}^2}{3Mf^2}\frac{1}{\beta
  L^3}\sum_{n_\mu\neq 0}
  \left(\frac{i}{\omega+\no+i\Delta}\right)^2
  \frac{(\nv)^4}{(\no)^2+(\nv)^2+m_\mathrm{eff}^2}\nn\\
  &=&-\frac{2g_{N\Delta}^2}{3Mf^2}{\mathbb B}(\Delta,m_\mathrm{eff}).
\end{eqnarray}
The functions ${\mathbb A}, {\mathbb B}, {\mathbb C}$ and ${\mathbb D}$ are
calculated in Appendix B. We reduced them to rapidly converging sums for
non-zero values of $m_\mathrm{eff}$ and $\Delta$, but no analytic form is
available. A {\it Mathematica} notebook computing these functions is available
from the authors' website~\footnote{http://nta0.lbl.gov/\~{}bedaque/index.html
  or http://ph.tum.de/\~{}hgrie}.  In the $\eps$ expansion, the contributions
coming from the finite values of $m$ and $\Delta$ appear only at order
$\eps^5$ in the loop diagrams, so we should take for these $m=\Delta=0$ at the
order $\eps^4$ we are working. In this case, a simple form for the nucleon
mass-shift is available:
\begin{equation}\label{M_strict}
  \delta M^{(3+4)} = \frac{1}{f^2L^3}\left(
  \frac{3g_A^2}{8}+\frac{2g_{N\Delta}^2}{3} \right) \left(
  1-\frac{\tau(\beta/L)}{ML} \right)-\frac{3\bar
  c_2}{f^2L^4}\left(\tau(\beta/L)-\frac{L}{\beta}\right)+ \frac{3 c_3}{f^2
  \beta L^3} -2m^2 c_1\left( \frac{X'(m^2f^2 \beta L^3/2)}{X(m^2f^2 \beta
    L^3/2)}-2 \right)\label{strict_epsilon}\;\;,
\end{equation}
where
\begin{equation}
  \gamma_0=\frac{1}{\pi^2}\sum_{\vec{j}\neq 0}\frac{1}{j^4}\approx
  1.675\;\;,\;\;
  \tau(x) =\frac{\gamma_0}{2}-\sum_{\vec{j}\neq 0}\frac{2\pi j}{e^{2\pi j x
  }-1},
\end{equation}
with $j=|\vec{j}|$.  The function $\tau(x)$ is plotted in Fig.~\ref{fig:tau}.

For not-so-small boxes satisfying $mL\sim \beta\Delta\agt 1$, the $p$
expansion applies. The two leading orders in the expansion of the nucleon mass
in the $p$-r{\'e}gime are very similar to the ones in the $\eps$-r{\'e}gime.
The differences are: i) leading and next-to-leading order are switched as the
quark mass insertion proportional to $c_1$ is the leading ($p^2$) order
contribution while the diagrams in Fig.~\ref{fig:lo} are the next-to-leading
($p^3$) contribution (terms proportional to $c_2,c_3$ are even higher, namely
$p^4$); and ii) the non-zero value of $m$ and $\Delta$ should be kept in the
diagrams. For this reason, if we keep the pion mass in our calculations, which
in the $\eps$-r{\'e}gime is a sub-leading ($\eps^5 $) effect, our expressions
will be valid in both r{\'e}gimes and, in particular, in the intermediate
region $(L,\beta)\cong 1/m$. This way, they also include some, but not all
${\mathcal O}(\eps^5)$ pieces~\footnote{Our results are also valid in the limit
  $\beta\to\infty$ as long as this limit is taken at fixed $m$.  If $m^2\beta$
  is kept fixed instead with $s=m^2\beta f^2L^3/2\sim 1$, the mass term does
  not prevent the ($\vec{n}=0, n_0\neq 0$) modes to have large fluctuations
  and they become non-perturbative. This is the $\delta$ r{\'e}gime discussed
  first in Ref.~\cite{leutwyler_delta}.}. Furthermore, since the pion mass in
the $\eps$-r{\'e}gime has a correction of order $\eps^0$ coming from the
integration over the zero-mode (Eq.(\ref{meff})), we use $m_\mathrm{eff}$ in
the one-loop diagrams.

\section{Euclidean time and the correct analytic continuation}

There is a subtlety in computing the nucleon mass using the combination of the
finite-temperature imaginary time and heavy baryon formalisms we used. To see
that, consider the derivation of the heavy baryon Lagrangean. One starts from
the relativistic nucleon field $\psi$ and performs a field redefinition which
reads in Euclidean space
\begin{equation}
  \psi(\tau,\vec{r})=e^{-M\tau}(N(\tau,\vec{r})+H(\tau,\vec{r})),
\end{equation} 
where $M$ is the heavy nucleon mass and $N$ and $H$ are the nucleon and
(anti)-nucleon fields satisfying $\gamma^0 N=N, \gamma^0 H=-H$.  An
``on-shell" $\psi$ field has a fast variation with time ($\partial_0 \psi \sim
M$), while an ``on-shell" $N$ satisfies $\partial_0 N\sim 0$. The Lagrangean in
terms of these new variables is
\begin{equation}
  \bar\psi(\partial_0\gamma_0+M)\psi \to N^\dagger(\partial_0+\cdots) N +
H^\dagger(\partial_0-2M+\cdots) H+\cdots .
\end{equation}
The ``heavy" field $H$ can then be integrated out and we are left with the
usual heavy baryon Lagrangean. Notice that the anti-periodic boundary
condition in the time direction for the relativistic field implies a different
boundary condition for the heavy-nucleon field
\begin{equation}
  \psi(\beta,\vec{r})=-\psi(0,\vec{r}) \Rightarrow
   N(\beta,\vec{r})=-e^{\beta M}N(0,\vec{r}).\label{bc}
\end{equation}
Therefore, the field $N$ has the Fourier decomposition
\begin{equation}
  N(\tau,\vec{r}) = \sum_{n_0}e^{-i(\frac{\pi(2n_0+1)}{\beta}+i
  M)\tau}N(n_0,\vec{r})\;\;.
\end{equation}
The correlators of the field $N$ are defined only at shifted values of
(imaginary) frequency, namely at $\omega=\pi(2n+1)/\beta+iM$.

Consider now, as an example, the computation of the first diagram in
Fig.~\ref{fig:lo}. For simplicity, we consider the infinite spatial volume
limit.  As shown in Appendix C, the sum over $n$ can be performed resulting up
to constants in
\begin{eqnarray}
  {\mathcal G}(\omega) &=& \frac{1}{\beta}\sum_n \int
\frac{d^3k}{(2\pi)^3}\vec{k}^2
\frac{i}{\frac{2\pi n}{\beta}+\omega+i\Delta}
\frac{1}{(\frac{2\pi n}{\beta})^2+\omega_k^2}\nn\\
&=&\int
\frac{dk^4}{(2\pi)^4}\frac{1}{k_0+\omega+i\Delta}
\frac{\vec{k}^2}{k_0^2+\omega_k^2}
-i\int
\frac{dk^3}{(2\pi)^3}\frac{\vec{k}^2}{(\omega+i\Delta)^2+\omega_k^2}
\frac{1}{e^{\beta(\Delta-i\omega)}-1}\nn\\
&\ \ &+\int
\frac{dk^3}{(2\pi)^3}\frac{\vec{k}^2}{(\omega+i\Delta)^2+\omega_k^2}
\frac{\omega+i\Delta}{\omega_k} \frac{1}{e^{\beta\omega_k}-1}.
\end{eqnarray}
where $\omega$ is the external energy and $\omega_k^2=\vec{k}^2+m^2$.  We now
substitute $\omega$ from above into the second term,
\begin{equation}
  \label{eq:tozero}
  \frac{1}{e^{\beta(\Delta-i\omega)}-1} =
 -\frac{1}{e^{\beta(M+\Delta)}+1}\approx 0,
\end{equation} 
leading to the correct statistics for fermionic ensembles.  Physically, that
we neglect these fluctuations just mirrors the fact that finite-temperature
fluctuations of heavy particles are much smaller than those of light ones for
temperatures $\beta M\gg 1$ at which the heavy-baryon formalism applies.
Therefore, we drop this term and arrive at
\begin{equation}
  {\mathcal G}(\omega) = \int
\frac{dk^4}{(2\pi)^4}\frac{1}{k_0+\omega+i\Delta}
\frac{\vec{k}^2}{k_0^2+\omega_k^2}
+\int
\frac{dk^3}{(2\pi)^3}\frac{\vec{k}^2}{(\omega+i\Delta)^2+\omega_k^2}
\frac{\omega+i\Delta}{\omega_k} \frac{1}{e^{\beta\omega_k}-1}.
\end{equation}
The nucleon propagator at any value of the external energy (including real
values) can be obtained from the expression above by analytically continuing
in $\omega$. In particular, the value determining the mass is obtained for
$\omega=0$.
 
Clearly, this procedure seems arbitrary for two reasons. First, it seems to
depend on the order between setting $\omega$ to $\omega=\pi(2n+1)/\beta+iM$
and analytically continuing to $\omega=0$. Second, the knowledge of the
propagator at discrete values of the frequency is not, in general, enough to
determine the propagator on the whole complex plane. One could, for instance,
have maintained $e^{-i\beta\omega}$ instead of substituting it by $-e^{\beta
  M}$ and using $e^{-2\pi in}=1$.  Fortunately, the analytic continuation is
unique for functions vanishing at infinity at least like
$1/|\omega|$~\cite{abrikosov}, as in the case at hand. Still, to confirm that
we have picked the correct analytic continuation, we repeat this calculation
in Appendix C without using the heavy baryon formalism in another method to
compute finite-temperature corrections which does not require an analytic
continuation to the real axis, namely the ``real time formalism".
 
\section{Numerical examples and discussion}

We now present some numerical examples in order to explore the convergence of
the $\eps$-expansion and to discuss how it can be used in the one baryon
sector. The leading order result depends on two low energy constants
$g_A=1.267$ and $g_{N\Delta}$, as well as on the masses and mass splittings
$m$, $M$ and $\Delta$, whose experimental values are reasonably well known. At
next-to-leading order, the constants $c_1, c_2$ and $c_3$ appear. They are
determined experimentally through the analysis of pion-nucleon
scattering. As they are most sensitive to the isoscalar part of the amplitude,
where different phase shift analyses disagree, large uncertainties exist in
their determination.
 
When considering very low energy observables, as we are here, the inclusion of
the $\Delta$ as explicit degree of freedom is optional. Let us first discuss
the case where the $\Delta$ is included. In this case, a determination of the
low energy constants was made by comparing calculations to the
pion-nucleon phase shift data \cite{fettes_delta}. In this work, different
fits were discussed using two different phase shift analyses, and also
including information about the $\sigma$-term. The values of $c_1$ and $c_3$
are more stable among different fits, while $c_2$ varies much more.

Eq.(\ref{strict_epsilon}) shows however that for a certain value of the ratio
$\beta/L\approx 1.22262$, $\tau(1.22262)=1/1.22262$ and the dependence on
$c_2$ disappears. Since $c_1$ and $g_{N\Delta}$ are much better determined,
one might use the mass-shifts measured around this ratio on the lattice to
determine $c_3$. The $c_2$-contribution is generically negligible for $\beta/L\sim[1\dots1.7]$.

\begin{figure}[!htbp]
\begin{minipage}[t]{3.5cm}
  \centerline{{\epsfxsize=2.6in \epsfbox{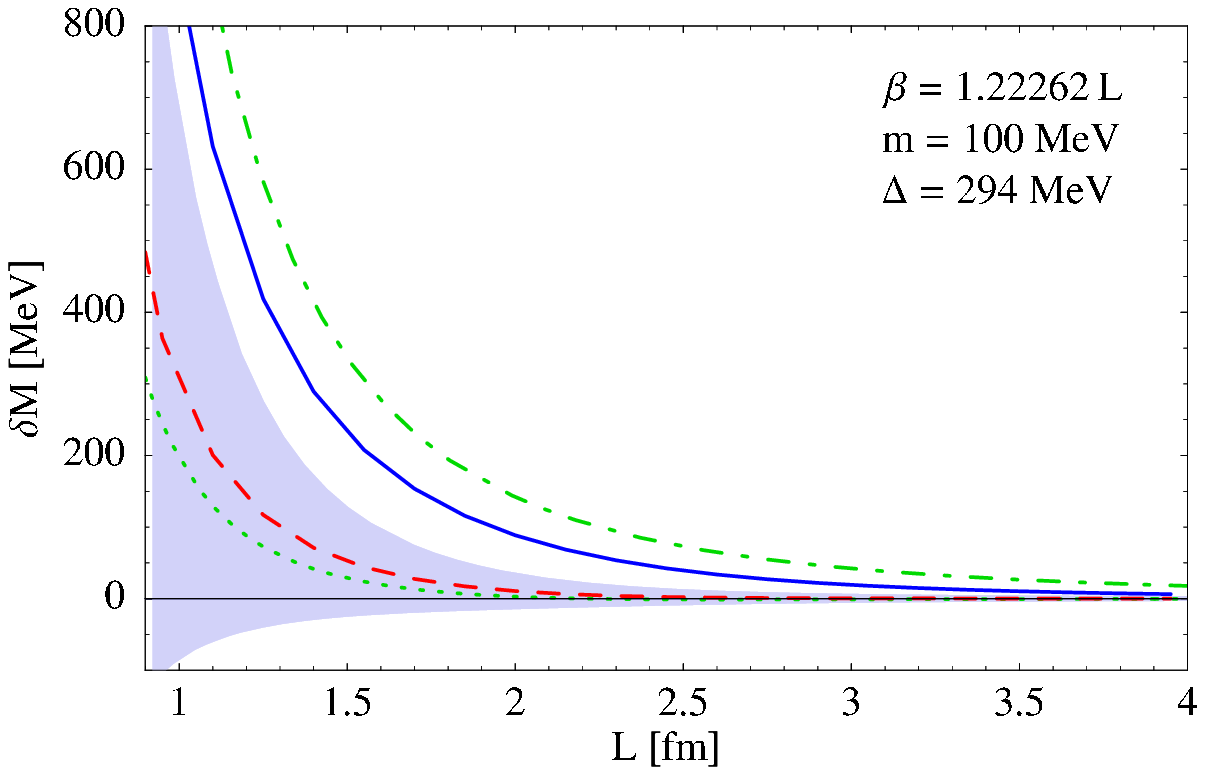}}}
\end{minipage}
\hskip 150pt
\begin{minipage}[t]{3.5cm}
  \centerline{{\epsfxsize=2.6in \epsfbox{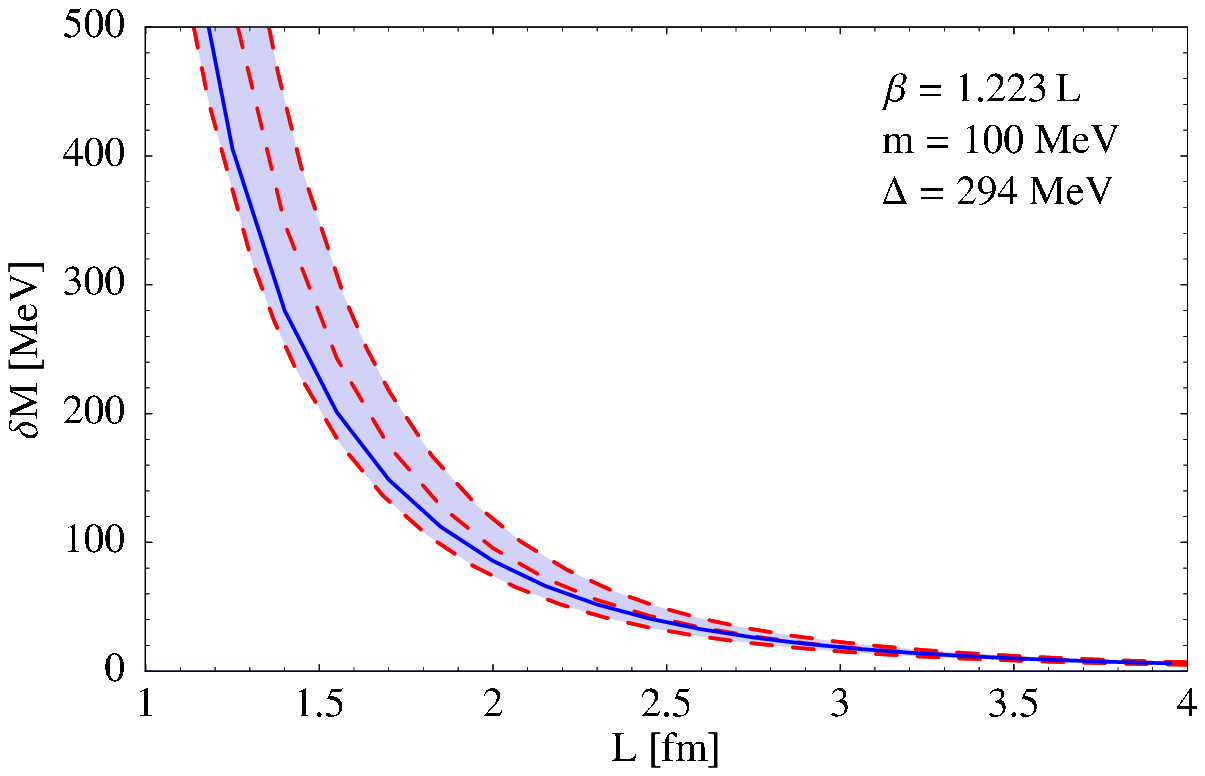}}}
\end{minipage}
\caption{\label{fig:m1}Left: Finite-volume mass-shift of the
  nucleon in the theory with explicit $\Delta$ degrees of freedom in MeV as
  function of $L$ [fm] with the central values of the parameter set in
  Eq.(\ref{fit2dagger}).  Leading order with full expansion (blue solid line);
  and using $m=\Delta=0$ Eq.(\ref{M_strict}) (green dash-dotted).
  Next-to-leading order {\it correction} in the full expansion (red dashed);
  and from Eq.(\ref{M_strict}) (green dotted). The gray zone shows the
  variation of the mass shift as $c_3$ varies in the range given in
  Eq.(\ref{fit2dagger}), with the upper limit corresponding to $c_3=1.8$/GeV.
  Right: Total mass-shift at leading order (blue solid line) and at leading +
  next-to-leading order for $c_3=0.06$/GeV, $0.93$/GeV and $1.8$/GeV from top
  to bottom (red dashed). The parameters $\beta=1.22262 L$, $m=100$ MeV and
  $\Delta=294$ MeV are the same for both figures.}
\noindent
\end{figure}

In Fig.~\ref{fig:m1}, we present results using the parameter set (fit
$2^\dagger$ of Table 4 in \cite{fettes_delta}) : \bea\label{fit2dagger}
g_{N\Delta}&=& 1.00\pm0.08,\nn\\
c_1&=&-0.35\pm 0.09/{\rm GeV},\nn\\
c_2&=&-1.49\pm 0.67/{\rm GeV},\nn\\
c_3&=&0.93\pm 0.87/{\rm GeV}.  \eea The errors quoted come from the fit and
under-estimate the uncertainty in the constants from higher-order corrections.
In the left panel of Fig.~\ref{fig:m1}, we show the leading contribution and
its next-to-leading order correction to the mass shift computed both using
Eq.(\ref{strict_epsilon}) and the full formula with finite $m_{eff}$ and
$\Delta$. The expansion in $m/(4\pi f), \Delta/(4\pi f)$ seems to converge for
the low-energy constants $c_i$ in the given range, except for those close to
the upper limit and for $L$ smaller than about $1.5$ fm, where one approaches
the breakdown scale: $L/(2\pi)\approx1/(4\pi f)$.  The right panel
displays the total mass shift up to second order for different values of $c_3$
between the minimum and maximum values suggested by Eq.(\ref{fit2dagger}). We
notice that the cancellation of the $c_2$-contribution for $\beta=1.22262 L$
works very well even for non-zero pion masses, with its contribution to the
mass shift never exceeding $5$ MeV even for $m=300$ MeV.

\begin{figure}[!htbp]
  \centerline{{\epsfxsize=2.6in \epsfbox{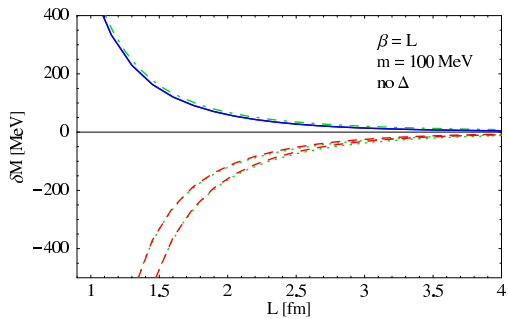}}}
\noindent
\caption{\label{fig:m5}Example of the finite-volume mass-shift of the
  nucleon at leading order (solid line) and the next-to-leading order {\it
    correction} (dashed lines) in MeV as function of $L $ [fm] in the theory
  without explicit $\Delta$ degrees of freedom. The lower dashed curve uses
  the central values of the parameter set in Eq.(\ref{buettiker}), the lower
  one changes only $c_3=-3.4$/GeV following~\cite{rent}. The pion mass is
  $m=100$ MeV, and $\beta=L$. Dot-dashed (dotted): LO (NLO) for $m=\Delta=0$.}
\end{figure}

In a effective theory without explicit $\Delta$, its large r{\^o}le in the
pion-nucleon interaction is absorbed by the couplings $c_2$ and $c_3$. In
fact, a simple tree level model of the $\Delta$ contribution gives a
contribution of $c_2=-c_3=g_A^2\Delta/(2(\Delta^2-m^2))\approx 4/$GeV. These
values are somewhat larger than what is expected from naive dimensional
analysis arguments and puts the quark mass expansion in check. 
The values suggested by different fits \cite{cs, rent, fettes}
roughly agree with the $\Delta$ saturation estimate. We show in
Fig.~\ref{fig:m5}, as an example, the mass shift computed with the central
values of the parameter set
\bea\label{buettiker}
c_1&=&-0.81\pm 0.15/{\rm GeV},\nn\\
c_2&=&2.99\pm 0.77/{\rm GeV},\nn\\
c_3&=&-4.70\pm 0.95/{\rm GeV}
\eea
advocated in \cite{cs}, as well as for a somewhat smaller value of
$c_3=-3.4$/GeV found e.g.~in the partial wave analysis of nucleon-nucleon
scattering~\cite{rent}. The convergence is obviously poor in either case.
While the contributions from $c_2$ and $c_3$ can be made small for certain
ratios $\beta/L$, this cancellation depends sensitively on the particular
values chosen for $c_2$ and $c_3$ and is hence less useful for lattice
determinations. At the ratio $\beta/L=1.22262$, the $c_2$-contribution
disappears as before, but it is already negligible at $\beta=L$. Indeed, a
plot of the next-to-leading order correction with $\beta/L=1.22262$ differs
from Fig.~\ref{fig:m5} at most at the 5\%-level.

One might wonder why the expansion in the $\eps$-regime is so sensitive on the
low-energy constants and converges badly
in the case without $\Delta$, while it is well behaved in the infinite volume
limit. This seems to arise because $c_2$ and $c_3$ appear only at third order
in the $p$-expansion and are, consequently, poorly determined. In the
$\eps$-r\'egime, they contribute already at second order, so that the
uncertainty in their values is enhanced.

\section{Conclusions}

We have computed the nucleon mass in a finite box of size $\beta\times L^3$
satisfying $4\pi f \gg 2\pi/\beta, 2\pi/L \gg m$ ($\eps$-regime). Taking the
value of the low energy constants suggested by experiment, we find that the
expansion seems to converge for the values of the low energy constants allowed
by fits made to pion-nucleon scattering data, if the $\Delta(1232)$ is taken
into account as explicit degree of freedom.  We notice that a particular shape
of the box ($\beta\approx 1.22262 L$) eliminates the dependence on one of the
low energy constants ($c_2$) and suggests determining the value of the most
poorly known one ($c_3$) by a fit to the nucleon mass with a few different box
sizes.  We also discussed a subtle point involving the combined use of the
imaginary time and heavy baryon formalisms.

We close with the remark that it may seem strange to compute pion loops in
small boxes of sizes $\approx 1$ fm$^4$, since the momentum of the first
non-zero mode $p\approx 1.2$ GeV is well above the range of validity of Chiral
Perturbation Theory. An alternative to our procedure that is not subject to
this criticism would be to integrate out all modes but the zero-mode and
obtain an effective theory which is valid only for zero-momentum observables
like the mass computed here.  We point out, however, that in order to connect
the low-energy constants of this new theory with the parameters of the
original chiral Lagrangean, one has to perform a matching calculation that is
equivalent to the calculation presented in this paper.
\section{Acknowledgements}

We would like to thank W.~Detmold, Th.~R.~Hemmert, N.~Kaiser, U.-G.~Mei\3ner
and M.~Savage for discussions on the topic.  H.W.G.~and G.R.~are grateful to
the Nuclear Theory Group at Lawrence Berkeley National Laboratory for warm
hospitality and financial support during the first stages of this work, and
H.W.G.~to the Nuclear Theory Group at Los Alamos National Laboratory for the
same reason.  H.W.G.~is supported in part by the Bundesministerium f\"ur
Forschung und Technologie and by the Deutsche Forschungsgemeinschaft under
contract GR1887/2-2. This work was supported in part by the U.S.~Department of
Energy under Contract No.~DE-AC03-76SF00098 and W-7405-ENG-36.

\section{Appendix A: Conventions}

We collect here the definitions used in the construction of the chiral
Lagrangean.  Elements of $SU_L(2)\times SU_R(2)$ are parameterized by $(\xi_L,
\xi_R)$.  $N$ is a spin and isospin doublet with its indices not shown
explicitly. $\Delta^{iA}$ is a spin-isospin $3/2$-field, where the vector and
isovector indices $jA$ are explicitly shown, besides the implicit spin and
isospin indices. In addition, it satisfies
$\sigma^i\Delta^{iA}=\tau^A\Delta^{iA}=0$ so only $4$ spin and $4$ isospin
entries are independent, as expected for (iso)spin $3/2$ objects. The chiral
transformation rules are
\begin{eqnarray}
\xi_L&\to& L \xi_L \hinv\nn,\\
\xi_R&\to& R \xi_R \hinv\nn,\\
N&\to& hN\nn,\\
\Delta^{iA} &\to& \underbrace{\frac{1}{2}\tr(\hinv\tau^Ah\tau^B)}_{\Op^{AB}} h
\Delta^{iB},
\end{eqnarray}
where $\Op^{AB}$ is an orthogonal matrix which is determined by $h$ via
$\hinv\tau^A h=\Op^{AB}\tau^B$, where $h$ in turn depends on $\pi(x), L$ and
$R$. In addition, we define some objects with simple chiral transformation
rules
\begin{eqnarray}
\V_\mu&=&\frac{1}{2}(\xi_R^\dagger\partial_\mu\xi_R+
\xi_L^\dagger\partial_\mu\xi_L)\to
h\V_\mu\hinv+h\partial_\mu\hinv,\nonumber\\ 
\A_\mu&=&\frac{i}{2}(\xi_R^\dagger\partial_\mu\xi_R-
\xi_L^\dagger\partial_\mu\xi_L)\to h\A_\mu\hinv,\nonumber\\ 
D_\mu N&=&\partial_\mu N+\V_\mu N\to hD_\mu N\nn,\\
D_\mu \Delta^{iA} &=& \partial_\mu \Delta^{iA} + \V_\mu \Delta^{iA} +i
\eps^{ABC} \tr(\tau^B\V_\mu)  \Delta^{iC}\to \Op^{AB}h D_\mu\Delta^{iB}\nn,\\ 
D_\mu\A_\nu &=& \partial_\mu\A_\nu+[\V_\mu, \A_\nu]\to h D_\mu\A_\nu
\hinv\nn,\\ 
w^A_\mu &=& \tr(\tau^A\A_\mu)\to \Op^{AB}w^B_\mu\nn,\\
w^A_{\mu\nu} &=& \tr(\tau^AD_\mu\A_\mu)\to \Op^{AB}w^B_{\mu\nu}.
\end{eqnarray}


\section{Appendix B: Calculation of sums}

In this appendix, we drop for clarity the subscript of the effective pion mass
$m_\mathrm{eff}$ and denote it by $m$. The LO diagrams contain then the sum
\begin{eqnarray}
  {\mathbb A} (\Delta,m)&=& \frac{1}{\beta L^3}
  \sum_{n_\mu\neq 0}\frac{1}{\omega+\no+i\Delta}
  \frac{(\nv)^2}{(\no)^2+(\nv)^2+m^2}\nn\\
  &=& {\mathbb A}_0(\Delta,m) + {\mathbb A}_\beta(\Delta,m),
\end{eqnarray}
where $\omega=2\pi(k+1/2)/\beta$ is a discrete external energy and $k$ an
integer. The ultraviolet divergence of ${\mathbb A}$ is identical to the one
in the infinite-volume diagram and cancels in the difference between finite
and infinite volume masses. For this cancellation to occur, it is important to
use the same regulator in both calculations. In practice, we should therefore
define the sum above using dimensional regularization.  ${\mathbb A}_0 $ and
${\mathbb A}_\beta $ are the temperature independent and the finite
temperature parts:
\begin{eqnarray}
  {\mathbb A}_0(\Delta,m) &=&\int \frac{dk_0}{2\pi}\frac{1}{L^3}\sum_{\vec{n}} \frac{1}{k_0+\omega+i\Delta}\frac{(\nv)^2}{k_0^2+(\nv)^2+m^2}\nn,\\
  {\mathbb A}_\beta(\Delta,m) &=& \frac{i}{L^3}\sum_{\vec{n}}
  \frac{\Delta-i\omega}{\omega_n^2-(\Delta-i\omega)^2}\frac{(\nv)^2}{\omega_n}
  \frac{1}{e^{\beta\omega_n}-1}- \frac{i}{L^3}\sum_{\vec{n}}
  \frac{(\nv)^2}{\omega_n^2-(\Delta-i\omega)^2}\frac{1}{e^{\beta(\Delta-i\omega)}-1},
\end{eqnarray}
where $\omega_n^2=\nv^2+m^2$. We used the formula
\begin{equation}
  \label{finiteTtrick} \frac{1}{\beta}\sum_{n} f(\frac{2\pi n}{\beta}) =
\int_{-\infty}^\infty \frac{dz}{2\pi} f(z) - i \text{Res}
(\frac{f(z)}{e^{i\beta z}-1})|_{\rm lower plane} + i \text{Res}
(\frac{f(z)}{e^{-i\beta z}-1})|_{\rm upper plane},
\end{equation}
which holds if $f(z)$ has no poles on the real axis. We substitute
$1/(e^{\beta(\Delta-i\omega)}-1)$ by $-1/(e^{\beta(\Delta+M)}+1)\approx 0$ as
in Eq.~(\ref{eq:tozero}).  The zero-temperature part can be computed with the
help of the relation \cite{elizalde}
\begin{eqnarray}
  \frac{1}{L^3}\sum_{\vec{n}} \frac{(\nv)^{2m}}{(\nv)^2+x^2} &=& \frac{1}{L^3}
\int d^3k \frac{k^{2m}}{k^2+x^2} \sum_{\vec{n}} \delta(\vec{k}-\frac{2\pi
  \vec{n}}{L}) = \int \frac{d^3k}{(2\pi)^3} \frac{k^{2m}}{k^2+x^2}
\underbrace{\sum_{\vec{n}} \delta(\frac{\vec{k}L}{2\pi}-\vec{n})
}_{\sum_{\vec{j}}e^{i L \vec{k}\cdot\vec{j}}}\nn\\
&=& \int \frac{d^3k}{(2\pi)^3} \frac{k^{2m}}{k^2+x^2} +
\int \frac{d^3k}{(2\pi)^3} \sum_{\vec{j}\neq 0}\frac{k^{2m}}{k^2+x^2}
e^{i L \vec{k}\cdot\vec{j}}\nn\\
&=& \int \frac{d^3k}{(2\pi)^3} \frac{k^{2m}}{k^2+x^2} + \sum_{\vec{j}\neq 0}
\frac{1}{2\pi^2L}\int_0^\infty dk \frac{k^{(2m+1)}}{k^2+x^2}
\frac{\sin(jkL)}{j}\nn\\
&=& \int \frac{d^3k}{(2\pi)^3} \frac{k^{2m}}{k^2+x^2} +
\frac{\left(-x^2\right)^m}{4\pi
  L}\sum_{\vec{j}\neq 0}\frac{e^{-jxL}}{j},\label{eq:poisson}
\end{eqnarray}
where $m$ is a positive integer.
Applying this relation to ${\mathbb A}_0$ yields
\begin{equation}
  \delta{\mathbb A}_0 = {\mathbb
  A}_0-{\mathbb A}(\beta\to\infty, L\to\infty) = \frac{i\Delta^2}{4\pi^2 L}
\sum_{\vec{j}\neq 0} \frac{1}{j} \underbrace{\frac{1}{\Delta} \int _0^\infty
  dk_0 \frac{k_0^2+m^2}{k_0^2+\Delta^2}e^{-jL\sqrt{k_0^2+m^2}}
}_{g(jL,m,\Delta)}.\nn
\end{equation}
Asymptotically, the sum over $j$ converges because
\begin{eqnarray}
  g(jL,m,\Delta) &\stackrel{j\to \infty}{\rightarrow}&
\frac{m^{5/2}}{\Delta^3}
\sqrt{\frac{\pi}{2jL}}e^{-jmL}+\cdots,\nn\\
g(jL,0,\Delta) &\stackrel{j\to \infty}{\rightarrow}&
\frac{2}{j^3L^3\Delta^3}+\cdots.
\end{eqnarray}
These asymptotic forms are also useful in the numerical evaluation of the sum
over $\vec{j}$.

We also need ${\mathbb A}_0$ evaluated at $\Delta=0$. We can obtain this limit
noticing that the integral defining $g(jL,m,\Delta)$ is nearly infrared
divergent when $\Delta\to 0$, and hence is dominated by small values of $k_0$.
The $\Delta\to 0$ limit of $g(jL,m,\Delta)$ is given by
\begin{equation}
  g(jL,m,\Delta)\stackrel{\Delta\to 0}{\longrightarrow} \frac{1}{\Delta}\int
  _0^\infty dk_0 \frac{k_0^2+m^2}{k_0^2+\Delta^2}e^{-jmL-jL\frac{k_0^2}{2m}} =
  \frac{\pi}{2}\frac{m^2}{\Delta^2}e^{-jmL}.
\end{equation}
Using this result,
\begin{equation}
  \delta{\mathbb A}_0 (\Delta=0,m)= \frac{im^2}{8\pi L}\sum_{\vec{j}\neq
  0}\frac{e^{-jmL}}{j}.
\end{equation}
The expression above agrees with that of Ref.~\cite{finite_volume}.  The limit
$m\to 0$ is found by noticing that for small values of $m$, the sum is
dominated by the large $j$ terms, which in turn can be approximated by an
integral
\begin{equation}
  \delta{\mathbb A}_0(\Delta=0,m) \stackrel{m\to 0}{\longrightarrow}
\frac{im^2}{8\pi L} 4\pi\int_0^\infty j e^{-jmL} \stackrel{m\to
  0}{\longrightarrow} \frac{i}{2L^3}.
\end{equation}
The double limit $\Delta\to 0, m\to 0$ can also be obtained in the opposite
order, and the result is the same:
\begin{eqnarray} \delta{\mathbb A}_0(\Delta,m=0)
  &=&
  \frac{i}{4\pi^2 L}  \sum_{\vec{j}\neq 0} \frac{1}{j}\Delta\int _0^\infty dk_0 \frac{k_0^2}{k_0^2+\Delta^2}e^{-jLk_0}\nn\\
  &=&\frac{i\Delta^2}{4\pi^2 L}  \sum_{\vec{j}\neq 0} \frac{1}{j} \underbrace{\left[\frac{1}{jL\Delta}-Ci(jL\Delta)\sin(jL\Delta)+Si(jL\Delta)\cos(jL\Delta)-\frac{\pi}{2}\cos(jL\Delta)\right]}_{g(jL,0,\Delta)} \nn\\
  &\stackrel{\Delta\to 0}{\rightarrow}& \frac{i\Delta^2}{4\pi^2 L} 4\pi \underbrace{\int_0^\infty dj j  g(jL,0,\Delta)}_{\frac{\pi}{2L^2\Delta^2}}=\frac{i}{2L^3}
\end{eqnarray}
The finite-temperature part converges very quickly:
\begin{eqnarray}
  \label{Abeta} {\mathbb A}_\beta &=& \frac{i}{L^3}\sum_{\vec{n}}
\frac{\Delta}{\omega_n^2-\Delta^2}\frac{(\nv)^2}{\omega_n}
\frac{1}{e^{\beta\omega_n}-1}\nn\\
&\stackrel{\Delta\to 0}{\longrightarrow}&0.
\end{eqnarray}
The second sum we need is
\begin{equation}
  {\mathbb B}(\Delta,m) = \frac{1}{\beta L^3}
  \sum_{n_\mu\neq
    0}\left(\frac{1}{\no+i\Delta}\right)^2\frac{(\nv)^4}{(\no)^2+(\nv)^2+m^2}.
\end{equation}
We use Eq.(\ref{finiteTtrick}) to separate it into a temperature-independent
(${\mathbb B}_0$) and a temperature-dependent part (${\mathbb B}_\beta$). The
first one is with 
Eq.~(\ref{eq:poisson})
\begin{equation}
  {\mathbb B}_0 = \int\frac{d^4k}{(2\pi)^4}
\frac{1}{(k_0+i\Delta)^2}\frac{\vec{k}^4}{k_0^2+\vec{k}^2+m^2} +
\frac{1}{\pi^2 L}\sum_{\vec{j}\neq 0} \frac{1}{j}
\int\frac{dk_0}{2\pi}\frac{1}{(k_0+i\Delta)^2}\int_0^\infty
dk\frac{k^{5-\eps}}{k_0^2+k^2+m^2}\sin(jkL) \;\;,
\end{equation}
\begin{equation}
  \delta{\mathbb B}_0 = {\mathbb B}_0 - {\mathbb B}(\beta\to\infty,
L\to\infty) =\frac{1}{4\pi^2 L}\sum_{\vec{j}\neq 0} \frac{1}{j} \underbrace{
  \int_0^\infty dk_0 \frac{k_0^2-\Delta^2}{(k_0^2+\Delta^2)^2}(k_0^2+m^2)^2
  e^{-jL\sqrt{k_0^2+m^2}} }_{\Delta^3 h(jL,m,\Delta)}.\label{Bh}
\end{equation}
The sums over $j$ converge, given the asymptotic behaviors
\begin{eqnarray}
  h( jL,m,\Delta)&\stackrel{j\to\infty}{\longrightarrow}& \frac{\sqrt{\pi
    m}}{4(jL)^{3/2}}\frac{e^{-jmL}}{\Delta^3}
(m+jL(\Delta^2-2m^2)),\nn\\
h(jL,m,0)&\stackrel{j\to \infty}{\longrightarrow}&
-\frac{12}{j^5L^5\Delta^5}+\cdots.\label{h_asym}
\end{eqnarray}
Eq.(\ref{h_asym}) can be obtained from the integral representation above
noticing that, for large $j$, the integral is dominated by small values of
$k_0$. These relations show that the sum in Eq.(\ref{Bh}) converges (quickly).

We also need the value of $\delta{\mathbb B}$ at $\Delta=0$, where
$h(j,L,m,\Delta)$ is {\it apparently} infrared divergent, but this limit is
actually finite. It is most easily obtained by continuing the $k_0$ integral
to $1+\eps$ dimensions and taking the $\eps\to 0$ limit at the end, with $K_n$
again a modified Bessel function:
\begin{eqnarray}
  \label{eq:B0Delta0}
  \delta{\mathbb B}_0(\Delta=0,m)&=&
  \frac{1}{4\pi^2L}\sum\limits_{\vec{j}\not=0}
  \frac{1}{j}\int\limits_0^\infty d^{1+\epsilon}k_0
  \frac{(k_0^2+m^2)^2}{k_0^2}e^{-jL\sqrt{k_0^2+m^2}}\nonumber\\
  &=& \frac{m^3}{4\pi^2L}\sum\limits_{\vec{j}\not=0}
  \frac{1}{j}\int\limits_1^\infty dx\frac{x^5}{(x^2-1)^\frac{3-\epsilon}{2}}
  e^{-jLm x}\nonumber\\
  &=& \frac{m^3}{4\pi^2L}\sum\limits_{\vec{j}\not=0}
  \frac{1}{j}\frac{\partial^4}{\partial(jLm)^4}
  \int\limits_1^\infty dx \frac{x}{(x^2-1)^\frac{3-\epsilon}{2}}
  e^{-jLm x}\nonumber\\
  &=&\frac{m}{4\pi^2L^3}\sum\limits_{\vec{j}\not=0}
  \frac{1}{j^3} \left[(jLm-(jLm)^3)K_0(jLm)+2(1+(jLm)^2)
    K_1(jLm)\right]\;\;,
\end{eqnarray}
We can further take the limit $m\to 0$:
\begin{equation}
  \delta{\mathbb B}_0(\Delta=0,m\to
0)=\frac{1}{2\pi^2L^4}\sum\limits_{\vec{j}\not=0}\frac{1}{j^4}=\frac{\gamma_0}{2L^4}.
\end{equation}
To take the double limit in the opposite order leads -- not surprisingly -- to
the same result:
\begin{eqnarray}
  \delta{\mathbb B}_0(\Delta,m=0)&=&
\frac{1}{4\pi^2L}\sum_{\vec{j}\neq 0} \frac{1}{j} \underbrace{ \int_0^\infty
  dk_0k_0^4\frac{k_0^2-\Delta^2}{(k_0^2+\Delta^2)^2}e^{-jL|k_0|}
}_{\Delta^3 h(jL,0,\Delta)} \nn\\
&=& \frac{\Delta^3}{8\pi^2L} \sum\limits_{\vec{j}\not= 0}\frac{1}{j}\left[
\frac{4-6(jL\Delta)^2}{(jL\Delta)^3}+2 Ci(jL\Delta) \left[ jL\Delta 
\cos(jL\Delta)+4\sin(jL\Delta)\right]\right. \nn\\
&& \left. \phantom{\frac{\Delta^3}{8\pi^2L} \sum\limits_{\vec{j}\not= 0}\frac{1}{j}
\frac{4-6(jL\Delta)^2}{(jL\Delta)^3}}
 +2 Si(jL\Delta) \left[ jL\Delta 
\sin(jL\Delta)-4\cos(jL\Delta)\right]\right]\nn\\
&\stackrel{\Delta\to 0}{\longrightarrow}&
\frac{1}{2\pi^2 L^4 }\sum_{\vec{j}\neq 0} \frac{1}{j^4}= \frac{\gamma_0}{2L^4}.\label{Bfinal}
\end{eqnarray}
After performing the correct analytic continuation, the temperature-dependent
part ${\mathbb B}_\beta$ is
\begin{equation}
  \label{Bbeta} {\mathbb B}_\beta = -\frac{1}{L^3}\sum_{\vec{n}}
\frac{\omega_n^2+\Delta^2}{(\omega_n^2-\Delta^2)^2}\frac{(\nv)^4}{\omega_n}
\frac{1}{e^{\beta\omega_n}-1}.
\end{equation}

Finally, we use similar steps for ${\mathbb C}$ and ${\mathbb D}$:
\begin{eqnarray}
  \label{eq:C}
  {\mathbb C}(m)&=&\frac{1}{\beta L^3}\sum\limits_{n_\mu\not=0}
  \frac{(2\pi n_0/\beta)^2}{(2\pi n_0/\beta)^2+(2\pi \vec{n}/L)^2+m^2}\nn\\
  &=&\int\frac{d^4k}{(2\pi)^4}\;\frac{k_0^2}{k_0^2+\vec{k}^2+m^2}
  +\frac{m^2}{(2\pi)^2 L^2}\sum\limits_{\vec{j}\not=0} \frac{K_2(jmL)}{j^2}
  -\frac{1}{L^3}\sum\limits_{\vec{n}} \frac{\omega_n}{e^{\beta\omega_n}-1}
  \;\;.
\end{eqnarray}
\begin{eqnarray}
  \label{eq:D}
  {\mathbb D}(m)&=&\frac{1}{\beta L^3}\sum\limits_{n_\mu\not=0}
  \frac{(2\pi\vec{n}/L)^2}{(2\pi n_0/\beta)^2+(2\pi
    \vec{n}/L)^2+m^2}\nn\\
  &=&\int\frac{d^4k}{(2\pi)^4}\;\frac{\vec{k}^2}{k_0^2+\vec{k}^2+m^2}
 -\frac{m^3}{(2\pi)^2 L} \sum\limits_{\vec{j}\not=0}
  \frac{1}{j}\left(K_1(jmL)+\frac{K_2(j m L)}{j m L}\right)
  +\frac{1}{L^3}\sum\limits_{\vec{n}}
  \frac{(\nv)^2}{\omega_n}
    \frac{1}{e^{\beta\omega_n}-1}
     \;\;,
\end{eqnarray}
For $m\to 0$, one has to be careful with the mode $\vec{n}=0$:
\begin{eqnarray}
  {\mathbb C}(m=0)&=&\frac{\gamma_0}{2 L^4}-
  \frac{1}{L^4}\sum\limits_{\vec{n}\not=0}\frac{2\pi n}
  {e^{2\pi\frac{\beta}{L}n}-1}-\frac{1}{\beta L^3}=
  \frac{\tau(\beta/L)}{L^4}-\frac{1}{\beta L^3}\nn\\
  {\mathbb D}(m=0)&=&-\frac{\gamma_0}{2 L^4}
  +\frac{1}{L^4}\sum\limits_{\vec{n}\not=0}\frac{2\pi
    n}{e^{2\pi\frac{\beta}{L}n}-1}=-\frac{\tau(\beta/L)}{L^4}.
\end{eqnarray}
 
\section{Appendix C: Relativistic calculation, real time formalism and
  the correct analytic continuation}

In order to verify our procedure to compute the finite temperature corrections
of the nucleon mass, we now repeat the calculation of the simplest diagram by
dispensing of the simplifications due to the use of both the heavy-baryon and
the imaginary time formalisms, followed by analytic continuation to the real
axis.

The {\it real-time} finite temperature formalism (RTF) is another way of
(perturbatively) computing finite-temperature corrections. As opposed to the
more common {\it imaginary-time} formalism (ITF), it computes correlators
directly in real time and continuous frequencies. The Feynman rules are very
similar to the ones at zero temperature, except that the propagators contain
an additional term describing the influence of the thermal medium on the
propagation of the particles~\footnote{In diagrams with more than one loop,
  the RTF rules are a little more involved.}.  The pion propagator becomes
\begin{equation}
  iD(k)=\frac{i}{k_0^2-\vec{k}^2-m^2+i0}+2\pi
n_B(|k_0|)\delta(k_0^2-\vec{k}^2-m^2),
\end{equation}
where $n_B(|k_0|)=(e^{\beta|k_0|}-1)^{-1}$ is the bosonic distribution
function. The fermion propagator is
\begin{eqnarray}
iS(p) &=& (ip_\mu\gamma^\mu+M)\left(\frac{1}{p_0^2-\vec{p}^2-M^2+i0}-
  2\pi n_F(|p_0|)\delta(p_0^2-\vec{p}^2-M^2)\right)\label{RTF-prop}\nn\\
&\stackrel{p_0\to M+k_0, \vec{p}\to\vec{k}}{\cong}&
\frac{i}{k_0+i0}-2\pi n_F(|M+k_0|)\delta(k_0)\label{RTF-heavy-prop},
\end{eqnarray}
where $n_F(k_0)=(e^{|k_0|}+1)^{-1}$ is the Fermi distribution function.  The
physical origin of these extra terms is the Pauli blocking (in the case of
fermions) or stimulated emission (in the boson case) caused by the real,
on-shell particles present in the medium. In the fermionic case, for instance,
a state that is fully occupied ($n_F=1$) reverts the sign of the ``$i\eps$"
prescription and the fermion can propagate as a hole. Notice that the number
density of particles in the heavy baryon propagator is $n_F(|M+k_0|)$ (as
opposed to $n_F(|k_0|)$) and therefore exponentially small at all temperatures
$\beta M\gg 1$ where the effective theory applies. As an example, let us
compute the real part of the second diagram in Fig.~\ref{fig:lo}, for
notational simplicity in the infinite volume. Up to irrelevant constants,
\begin{eqnarray}
  i G(E) &=& \int\frac{d^4k}{(2\pi)^4}\vec{k}^2
((E+k_0)\gamma^0-\vec{k}\cdot\vec{\gamma}+M_\Delta) \left[
  \frac{i}{k_0^2-\vec{k}^2-m^2+i0}+2\pi n_B(|k_0|)\delta(k_0^2-\vec{k}^2-m^2)
\right]
\nn\\
& &\ \ \ \ \ \ \ \ \ \ \ \ \ \ \ \ \ \ \ \ \ \ \left[
  \frac{1}{(E+k_0)^2-\vec{k}^2-M_\Delta^2+i0}+2i\pi
  n_F(E+k_0)\delta((E+k_0)^2-\vec{k}^2-M_\Delta^2) \right].
\end{eqnarray}
Using the relation
\begin{equation}
  \frac{1}{x+i0}={\mathcal P}\left( \frac{1}{x} \right)-i\pi \delta(x),
\end{equation}
where ${\mathcal P}$ stands for the principal value, the real part of
$G(\omega)$ is
\begin{eqnarray}
  {\mathcal Re} G(\omega) &=& \int \frac{d^3k}{(2\pi)^3}\vec{k}^2 \left[
  \frac{1+2n_B(\omega_k)}{2\omega_k} \left(
    \frac{(E+\omega_k)\gamma^0-\vec{k}\cdot\vec{\gamma}+M_\Delta}{
      (E+\omega_k)^2-\vec{k}^2-M_\Delta^2}+
    \frac{(E-\omega_k)\gamma^0-\vec{k}\cdot\vec{\gamma}+M_\Delta}{
      (E-\omega_k)^2-\vec{k}^2-M_\Delta^2}
  \right)
\right.\nn\\
&&+ 
  \frac{1-2n_F(\sqrt{\vec{k}^2+M_\Delta^2})}{2\sqrt{\vec{k}^2+M_\Delta^2}}
  \left(-i\sqrt{\vec{k}^2+M_\Delta^2}\gamma^0-i\vec{k}\cdot
    \vec{\gamma}+M_\Delta\right)\nn\\
  &&\ \ \ \ \ \ \ \ \left.\times
  \left( \frac{1}{(-E+\sqrt{\vec{k}^2+M_\Delta^2})^2-\omega_k^2}+
    \frac{1}{(E+\sqrt{\vec{k}^2+M_\Delta^2})^2-\omega_k^2} \right) \right]
\nn\\
&\stackrel{E=M+\omega}{\approx}& \int \frac{d^3k}{(2\pi)^3}\;
  \vec{k}^2 {\mathcal P}\left( \frac{1}{(\Delta-\omega)^2-\omega_k^2} \right)
  \left(
  \frac{\omega-\Delta}{2\omega_k}(1+2n_B(\omega_k)) +
  \frac{1-2n_F(M+\Delta-\omega)}{2} \right).
\end{eqnarray}
This last result can also be obtained directly using the RTF with the heavy
baryon propagator in Eq.(\ref{RTF-heavy-prop}).

On the other hand, we compute a related quantity, namely the Euclidean time
Matsubara function ${\mathcal G}(\omega)$ defined for discrete imaginary
values of $E=-i\pi(2n+1)/\beta$, as
\begin{eqnarray}
  {\mathcal Re}{\mathcal G}(E)&=&\frac{1}{\beta}\int \frac{d^3k}{(2\pi)^3}\vec{k}^2
\frac{-i(E+\omega_n)\gamma^0-i\vec{k}\cdot\vec{\gamma}^2+M_\Delta}{(E+\omega_n)^2+\vec{k}^2+M_\Delta^2}
\frac{1}{\omega_n^2+\omega_k^2}\nn\\
&=&\int \frac{d^4k}{(2\pi)^4}\vec{k}^2
\frac{-i(E+k_0)\gamma^0-i\vec{k}\cdot\vec{\gamma}^2+M_\Delta}{(E+k_0)^2+\vec{k}^2+M_\Delta^2}
\frac{1}{k_0^2+\omega_k^2}\nn\\
&&-\int \frac{d^3k}{(2\pi)^3}\vec{k}^2 \frac{n_B(\omega_k)}{2\omega_k} \left[
  \frac{-i(E+i\omega_k)\gamma^0-i\vec{k}\cdot\vec{\gamma}^2+M_\Delta}{(E+i\omega_k)^2+\vec{k}^2+M_\Delta^2}+
  \frac{-i(E-i\omega_k)\gamma^0-i\vec{k}\cdot\vec{\gamma}^2+M_\Delta}{(E-i\omega_k)^2+\vec{k}^2+M_\Delta^2}
\right]
\nn\\
&&-\int \frac{d^3k}{(2\pi)^3}\vec{k}^2 \frac{1}{2\sqrt{\vec{k}^2+M_\Delta^2}}
\left[
  \frac{i\sqrt{\vec{k}^2+M_\Delta^2}\gamma^0-i\vec{k}\cdot\vec{\gamma}^2+M_\Delta}
  {(-E+i\sqrt{\vec{k}^2+M_\Delta^2})^2+\omega^2}
  \frac{1}{e^{\beta\sqrt{\vec{k}^2+M_\Delta^2}+i\beta E}-1}\right.\nn\\
&&\ \ \ \ \ \ \ \ \ \ \ \ \ \ \ \ \ \ \ \ \ \ \ \ \ \ \ \ \ \ \ +\left.
  \frac{-i\sqrt{\vec{k}^2+M_\Delta^2}\gamma^0-i\vec{k}\cdot\vec{\gamma}^2+M_\Delta}
  {(E+i\sqrt{\vec{k}^2+M_\Delta^2})^2+\omega^2}
  \frac{1}{e^{\beta\sqrt{\vec{k}^2+M_\Delta^2}-i\beta E}-1} \right]
\end{eqnarray}
We now substitute again $e^{\pm i\beta E}$ by $ -1$.  The real part of $G(E)$
is related to the Matsubara function ${\mathcal G}(\omega)$ by analytic
continuation: $G(E)=-{\mathcal G}(-iE+0)$~\cite{abrikosov} . The direct
calculation of $G(E)$ using the RTF and the indirect one through analytic
continuation from the ITF agree, and they also reproduce in the limit
$M\to\infty$ the calculation using both the ITF and the heavy baryon formalism
discussed in the main text and in Appendix B.

  

\begin{thebibliography}{99}
  
  \bibitem{finite_volume} S.~R.~Beane,
    [arXiv:hep-lat/0403015],
    S.~R.~Beane and M.~J.~Savage,
    [arXiv:hep-ph/0404131],
    M.~Procura, T.~R.~Hemmert and W.~Weise,
    Phys.\ Rev.\ D {\bf 69}, 034505 (2004) [arXiv:hep-lat/0309020],
    A.~Ali Khan {\it et al}, [arXiv:hep-lat/0312030].
    
  \bibitem{leutwyler_epsilon}
    J.~Gasser and H.~Leutwyler,
    Phys.\ Lett.\ B {\bf 188}, 477 (1987).
    
  \bibitem{epsilon_mesons} L.~Giusti, P.~Hernandez, M.~Laine, P.~Weisz and
    H.~Wittig,
    JHEP {\bf 0404}, 013 (2004) [arXiv:hep-lat/0402002],
    L.~Giusti, P.~Hernandez, M.~Laine, P.~Weisz and H.~Wittig,
    JHEP {\bf 0401}, 003 (2004) [arXiv:hep-lat/0312012],
    P.~Hernandez and M.~Laine,
    JHEP {\bf 0301}, 063 (2003) [arXiv:hep-lat/0212014],
    W.~Bietenholz, T.~Chiarappa, K.~Jansen, K.~I.~Nagai and S.~Shcheredin,
    JHEP {\bf 0402}, 023 (2004) [arXiv:hep-lat/0311012],
    
  \bibitem{leutwyler_delta} H.~Leutwyler, Phys. \ Lett. {\bf B}189, 197
    (1987).

  \bibitem{abrikosov} Abrikosov, Gorkov and Dzyaloshinski, Methods of Quantum
    Field Theory in Statistical Mechanis, Dover, 1975.
    
  \bibitem{fettes_delta}  N.~Fettes and U.~G.~Mei\3ner,
   Nucl.\ Phys.\ A \textbf{679} (2001), 629 [arXiv:hep-ph/0006299].
  
  \bibitem{cs} P.~B\"uttiker and U.~G.~Mei\3ner,
    Nucl.\ Phys.\ A {\bf 668} (2000) 97 [arXiv:hep-ph/9908247].
    
  \bibitem{rent} M.~C.~M.~Rentmeester, R.~G.~E.~Timmermans and J.~J.~de Swart,
    Phys.\ Rev.\ C {\bf 67} (2003) 044001 [arXiv:nucl-th/0302080].

  \bibitem{fettes} N.~Fettes, U.~G.~Mei\3ner and S.~Steininger,
    Nucl.\ Phys.\ A {\bf 640} (1998) 199 [arXiv:hep-ph/9803266].

  \bibitem{elizalde} E.~Elizalde, Comm. Math. Phys. 198,83 (1998).
\end{thebibliography}
\end{document}